\documentstyle[psfig]{l-aa}

\newcommand{\less}{\raisebox{-1.1mm}{$\stackrel{<}{\sim}$}}
\newcommand{\more}{\raisebox{-1.1mm}{$\stackrel{>}{\sim}$}}

\newcommand{\msol}{{M$_{\odot}$}}
\newcommand{\mdot}{{$\dot{M}$}}
\newcommand{\lsol}{{L$_{\odot}$}}

\newcommand{\ks}{km s$^{-1}$}
\newcommand{\msolyr}{M$_{\odot}$ yr$^{-1}$}

\include{psfig.sty}

\begin{document}


\thesaurus{06(08.03.4, 08.09.2: IRC~+10~216, 08.13.2, 08.16.4, 13.19.5)} 
\title{IRC +10 216 revisited II: the circumstellar CO shell
\thanks{The James Clerk Maxwell Telescope is operated by the
Joint Astronomy Centre on behalf of the Particle Physics and Astronomy
Research Council of the United Kingdom, 
the Netherlands Organization for Scientific Research,
and the National Research Council of Canada.}
}

\author{M.A.T. Groenewegen\inst{1} \and W.E.C.J. van der Veen\inst{2} 
\and H.E. Matthews\inst{3} 
}
\offprints{Martin Groenewegen (groen@mpa-garching.mpg.de)}
\institute {Max-Planck-Institut f\"ur Astrophysik, 
Karl-Schwarzschild-Stra{\ss}e 1, D-85740 Garching, Germany
\and
Department of Astronomy, Columbia University, 538 West 120th Street, 
New York, NY 10027, U.S.A.
\and
Joint Astronomy Centre, 660 N. A'oh\={o}k\={u} Place, University Park, Hilo, 
Hawaii 96720, U.S.A., and Herzberg Institute of Astrophysics, NRC of
Canada, 5071 W. Saanich Road, Victoria, B.C. V8X 4M6, Canada
}
\date{received 29 September 1997,  accepted 30 June 1998}

\maketitle

\begin{abstract}

$^{12}$CO and $^{13}$CO J = 6-5 observations of IRC~+10~216 with the
JCMT are presented.  A spherically symmetric radiative transfer code
is used to model these and other CO observations of the carbon star
IRC~+10~216/CW Leo. Compared to previously published model
calculations a much larger set of observational data is used as
constraints; on-source $^{12}$CO and $^{13}$CO J = 1-0 up to J = 6-5
and mapping data taken with various telescopes, most of which are
obtained from the literature.

The gas temperature in the envelope is calculated taking into account
heating and cooling. The heating by dust-gas collisions and various
other parameters (such as luminosity over distance squared) are
constrained from our previous modeling of the circumstellar dust shell.
Photoelectric heating is taken into account.

A grid of models is calculated with the following parameters:
luminosity (in the range 10~000 -- 30~000, in steps of 5~000 \lsol),
mass loss rate, dust-to-gas ratio, dust opacity and CO abundance.  For
each of the considered luminosities a good fit to the on-source data
can be obtained. A comparison with CO J = 1-0 data obtained from the
literature points towards a preferred luminosity of 10-15~000 \lsol.

Notwithstanding the overall good agreement, there remain
discrepancies. The different observed $^{12}$CO J = 3-2 observations
appear to be always larger than the model predictions. The observed
$^{13}$CO J = 6-5 is almost flat-topped, while the model gives a
slight double-peaked profile. There is a large discrepancy with the
single existing $^{12}$CO J = 7-6 observation.

The best fitting models (for each of the considered luminosities)
cannot accommodate the more extended emission seen in the mapping
data.  This is not due to an underestimate of the photoelectric
effect.  To fit the data for radial offsets $>$50\arcsec\ the mass
loss rate must be a factor of 5 higher in the outer envelope.  Because
the various sets of data for offsets \more 150\arcsec\ are not
consistent with each other it is unclear if the enhancement in the
mass loss rate extends beyond that radius.

Visibility curves are calculated for comparison with future
interferometric observations. These appear to be insensitive to
luminosity and mass loss variations but should be good tracers of the 
geometry of the CO shell.

A comparison is made between the mass loss rates and dust-to-gas
ratios derived from the CO modelling and those derived from our
previous dust modelling. To do this we make use of the
relation $\dot{M} \; v_{\infty} = {\tau}_{\rm F} \; \frac{L}{c} \;
\left(1 - \frac{1}{\Gamma}\right)$, which is valid for radiation
pressure driven outflows. The best agreement is obtained for the model
with 15~000 \lsol. This agrees well with the luminosity range
7~700-12~500 \lsol\ based on the period-luminosity relation for carbon
miras.

In summary, we conclude that the likely luminosity of IRC~+10~216 
is between 10~000 and 15~000 \lsol\ and that its distance is between
110 and 135 pc. The present-day mass loss rate is (1.5 $\pm$ 0.3)
$\times 10^{-5}$ \msolyr\ and the gas-to-dust ratio is a 700 $\pm$
100. The dust opacity at 60 $\mu$m is found to be of order of 250
cm$^2$gr$^{-1}$. The CO abundance is 1.1 $\times$ 10$^{-3}$ relative
to H$_2$.

\vspace{-2mm}

\vspace{-4mm}
\end{abstract}
\vspace{-4mm}

\section{Introduction}

IRC~+10~216 (= AFGL 1381 = CW Leo) is the best-studied carbon star,
because it is intrinsically bright and nearby.  One aspect of study
has traditionally been the properties of the circumstellar shell:
e.g., what is the mass loss rate, has the mass loss rate changed with
time, what kind of chemistry takes place, what is the geometry of the
shell?

In two recent papers we have made an extensive study of the
circumstellar dust shell of IRC~+10~216 (Groenewegen 1996, 1997,
hereafter G97). A much larger set of observational constraints was
used than in any previous study.  This allowed the derivation of the
present-day dust mass loss rate, the ratio between the luminosity at
maximum light divided by the distance squared, the dominant grain
size, the effective temperature and the inner dust radius.

In another paper we have mapped the 1.3 mm continuum emission around
the star (Groenewegen et al. 1997). It was detected out to 50\arcsec\
at a noise level of 3$\sigma$ or better. There is evidence for phases
of enhanced mass loss, but a quantitative analysis is hampered by the
significant contribution of molecular lines to the continuum emission,
and the uncertainty in the radial distribution of that contribution.

Another way to study the circumstellar shell is by observing the gas
component.  A good tracer of the (H$_2$) gas is the CO molecule, and
numerous CO observations exist of IRC~+10~216 (see the catalog of Loup
et al. (1993) for a compilation).  A fair number of detailed models
have been proposed to explain various CO observations. The pioneering
work was done by Kwan \& Hill (1977). For an assumed distance of 200
pc they deduced a mass loss rate of 2 $\times$ 10$^{-5}$ \msolyr, a CO
abundance relative to H$_2$ of about 8 $\times$ 10$^{-4}$ (both
uncertain to a factor of 2), and a CO/$^{13}$CO isotope ratio of 35
$\pm$ 7. In a second paper Kwan \& Linke (1982) deduced a mass loss
rate of 4 $\times$ 10$^{-5}$ \msolyr\ for a distance of 200 pc and a
CO abundance of 6 $\times$ 10$^{-4}$.  Sahai (1987) finds, based on
the modeling of high-J lines, that the temperature in the inner region
is much larger than that predicted by the Kwan \& Linke model, and
that the mass loss rate in the inner 6\arcsec\ region is 3.2 $\times$
10$^{-5}$ \msolyr\ and 50\% higher in the extended envelope, assuming
a distance of 300 pc and a CO abundance of 8 $\times$ 10$^{-4}$.
Truong-Bach et al. (1991) also find evidence for a hot inner core: a
mass loss rate of 2.5 $\times$ 10$^{-5}$ \msolyr\ in the inner
4.2\arcsec\ and a mass loss rate of 4 $\times$ 10$^{-5}$ \msolyr\ in
the outer region (for a distance of 200 pc).  Kastner (1992) found a
best fit to some CO data for a distance of 150 pc and a mass loss rate
of 2 $\times$ 10$^{-5}$ \msolyr.  
In a recent paper, Crosas \& Menten (1997) also presented J = 6-5
$^{12}$CO and $^{13}$CO data, and combined those with other data from
the literature, and modelled the data. In this respect the philosophy
behind that paper is similar to ours. The major differences with that
paper are that the effective absorption coefficient is taken into
account in a consistent way (see below and Sect.~7.2) and that the
parameter space is scanned consistently no a-priori fixing any
parameter. In Crosas \& Menten, the dust-to-gas ratio is {\it
fixed} at a value of 0.01.  They find a good fit using a distance of
150 pc, a mass loss rate of 3.25 $\times$ 10$^{-5}$ \msolyr, a CO
abundance of 6 $\times$ 10$^{-4}$, and a dust-gas momentum transfer
rate $Q$ of 0.025 for the inner region ($r \le 10^{16}$ cm), and 0.018
for the outer region.  They conclude there is no need for a hot inner core.

In all the model calculations mentioned above the effective absorption
coefficient $Q$ is assumed to be a free parameter. $Q$ describes the
momentum transfer efficiency between the gas and the dust. However, we
claim, and this is a key point in this paper, that effective
absorption coefficient $Q$ is not a free parameter but is related to
the properties of the dust. This had been realised before but it
requires the simultaneous fitting of the spectral energy distribution
(SED) and the CO data in order to estimate $Q$ in a self-consistent
way. This approach was taken in the fitting of two OH/IR stars in
Groenewegen (1994b) and is proposed here for IRC~+10~216.

The outline of the paper is as follows. In Sects. 2 and 3 the model is
briefly described. In Sect. 4 the $^{12}$CO and $^{13}$CO J = 6-5 
observations as well as the other CO data to which the models are
compared are presented. In Sect. 5 the results of the model
calculations are described. So-called visibility curves, to be
compared with future interferometric observations, are presented in
Sect. 6. The paper ends with discussion and conclusions in Sect. 7.

\section{The model}

\begin{table*}[th]
\caption[]{CO data used as constraints}
\begin{flushleft}
\begin{tabular}{llllll} \hline
Molecule & Transition  & Reference                 & Telescope & Beam width & Remark \\ 
         &             &                           &           & FWHM(\arcsec) &  \\ \hline
$^{12}$CO & 1-0        & Groenewegen et al. (1996) & IRAM  & 21.0& primary constraint \\
          &            & Truong-Bach et al. (1991) & IRAM  & 21.0 & map \\
          &            & Groenewegen \& Ludwig (1998) & IRAM  & 21.0 & map \\
          &            & Knapp \& Morris (1985)    & BTL   & 100 & \\
          &            & Huggins et al. (1988)     & NRAO  & 50 & map \\
          &            & Olofsson et al. (1993)    & SEST  & 45 & \\
          &            & Huggins et al. (1988)     &  OSO  & 34 & map \\
          & 2-1        & Groenewegen et al. (1996) & IRAM & 12.5 & primary constraint\\
          &            & Truong-Bach et al. (1991) & IRAM  & 12.5 & map \\
          &            & Groenewegen \& Ludwig (1998) & IRAM  & 12.5 & map \\
          &            & Huggins et al. (1988)     & NRAO & 31 & map \\
          &            & Olofsson et al. (1993)    & SEST & 23 & \\
          & 3-2        & Groenewegen et al. (1996) & JCMT & 14.3 &   primary constraint\\
          &            & Williams \& White  (1992) & JCMT & 15 & \\
          &            & Wang et al. (1994)        & CSO  & 21.4 &  \\
          & 4-3        & Williams \& White  (1992) & JCMT & 11.0 & primary constraint\\
          & 6-5        & this paper                 & JCMT & 55\% in
7.0\arcsec\ + 45\% in 18\arcsec\ beam & primary constraint\\
          & 7-6        & Wattenbach et al. (1988)  & UH 2.2m & 45 & \\
$^{13}$CO & 1-0        & Groenewegen et al. (1996) & IRAM & 21.0 & primary constraint\\
          &            & Knapp \& Chang (1985)     & BTL  & 105  & \\
          & 2-1        & Groenewegen et al. (1996) & IRAM & 12.5 & primary constraint\\
          & 3-2        & Groenewegen et al. (1996) & JCMT & 14.3 & primary constraint\\
          &            & Williams \& White  (1992) & JCMT & 15 & \\
          &            & Wang et al. (1994)        & CSO  & 20.3 &  \\
          & 6-5        & this paper                 & JCMT & 55\% in
7.0\arcsec\ + 45\% in 18\arcsec\ beam & primary constraint\\
\hline
\end{tabular}
\end{flushleft}

\end{table*}

The calculations are performed with the molecular line emission
radiative transfer model of Groenewegen (1994a; G94). This model is an
improvement over and extension of the code by Morris et al. (1985)
through the inclusion of a self-consistent determination of the gas
temperature. The model was previously applied to two OH/IR stars at
known distances (Groenewegen 1994b).

The model is fully described in G94 and only some essential details
are repeated here. The model assumes spherical symmetry and that the
Sobolev approximation is valid to calculate the radiative transfer
(this presumes that the local linewidth is much smaller than the
expansion velocity). The line profiles are calculated without the
Sobolev approximation using a ray-tracing method instead, assuming, in
the present paper, a thermal velocity of 0.5 \ks. The CO,
respectively $^{13}$CO or HCN, molecules are excited by: (1) collisions with
H$_2$ and He molecules, (2) interaction with the 2.8 K background
radiation, and (3) infrared radiation from a central blackbody of
temperature $T_{\rm BB}$ and radius $R_{\rm BB}$ which leads to
pumping from the v = 0 into the $\mbox{v = 1}$ vibrational state.


The heating rate by the photoelectric effect is
discussed in G94. The heating rate per hydrogen molecule is not simply
a constant as is often assumed, but the grain charge parameter is
calculated for each radius and optical depth effects are taken into
account. The photoelectric heating rate scales as ($G_0\, (Y/0.1)$),
with $G_0$ the UV flux of the diffuse interstellar medium, and $Y$ the
yield of dust grains. $G_0$ = 1 and $Y$ = 0.1 are used unless
otherwise noted.

The heating rate per unit volume caused by dust-gas collisions is 
(see G94 for details):


\begin{equation}
H_{\rm dg} \sim  n(H_2)^2 \;\, \frac{\Psi \; (1 + 4 \; 
f_{\rm He})^2}{a \;{\rho}_d} \; \left(\frac{L \; Q \;
v(r)}{\dot{M}}\right)^{3/2} 
\end{equation}
and the drift velocity in km s$^{-1}$ given by:
\begin{equation}
v_{\rm dr} = 
1.4293 \times 10^{-4} \; \left(\frac{L \; Q \; v(r)}{\dot{M}}\right)^{0.5}
\end{equation}
where ${\rho}_{\rm d}$ is the dust grain density, 
$a$ the grain size, 
$L$ the stellar luminosity in solar units,
$\dot{\rm M}$ the mass loss rate in \msol/yr, $Q$ the effective
absorption coefficient (defined below in Eq. 4), $v$(r) the gas velocity in
km s$^{-1}$, $f_{\rm He}$ the number ratio of helium to hydrogen (a
value of 0.1 is assumed) and $\Psi$ the dust-to-gas ratio.

The cooling rate is determined by molecular line cooling of $^{12}$CO,
$^{13}$CO, HCN, H$_2$ and adiabatic cooling. The H$_2$ cooling rate is
treated in an approximate way but is not very important (G94).  The
cooling rate due to $^{12}$CO, $^{13}$CO and HCN is calculated from
the molecular excitation code in an iterative process. For a given
temperature structure the level populations are calculated, which allows
the molecular cooling to be determined, and finally a new temperature
structure is derived by solving the energy balance equation (see G94).

\subsection{How does dust affect the  molecular excitation model?}

As mentioned before dust grain-gas collisions is the most important
heating mechanism and therefore the properties of the dust (size,
opacity, dust-to-gas ratio) influence the gas temperature and hence the
resulting CO line profiles.

In G97 the spectral energy distribution (SED) and visibility curves of
IRC~+10~216 were modelled using a dust radiative transfer (DRT) model
and constraints  were set to the dust properties. Table~2 lists for an
assumed luminosity, the derived mass loss rate and distance.

\begin{table}
\caption[]{Default parameters}
\begin{flushleft}
\begin{tabular}{lll} \hline
Luminosity & Distance$^{(1)}$  & ${\dot{M}}_{\rm def}$ $^{(1)}$ \\
 (\lsol)   &  (pc)     & 10$^{-5}$ \msolyr              \\ \hline
 10~000  & 110 & 1.80 \\
 15~000  & 135 & 2.20 \\
 20~000  & 156 & 2.54 \\
 25~000  & 174 & 2.84 \\
 30~000  & 191 & 3.11 \\
\hline
\end{tabular}
\end{flushleft}

\noindent
$^{(1)}$Parameters that follow from the dust modeling in G97. The mass
loss rate is based on $\Psi$ = 0.005, ${\kappa}_{60}$ = 68
cm$^2$gr$^{-1}$ and a dust expansion velocity of 17.5 \ks.

\end{table}

\begin{table*}[th]
\caption[]{Grid of models ran}
\begin{flushleft}
\begin{tabular}{lrrrrrrrrrr} \hline
    & $\Psi$& $\Psi$/1.2 & $\Psi$/1.25 & $\Psi$/1.5 & $\Psi$/1.67 &
$\Psi$/1.8 & $\Psi$/2  & $\Psi$/3  & $\Psi$/4 & $\Psi$/5\\ \hline
\multicolumn{2}{c}{L = 10~000 \lsol} & & & \\
        & & & & \\
${\dot{M}}_{\rm def} \times$ 1.0 &  9 &- &- &- &- &- & 10 & 10 & - & 9  \\
${\dot{M}}_{\rm def}$ / 1.2      &  - &- &- &- &- &- &- & 10 & {\bf  11}  & -  \\
${\dot{M}}_{\rm def}$ / 1.4      &  - &- &- &- &- &- &- & 13 & -  & - \\
${\dot{M}}_{\rm def}$ / 1.8      &  - &- &- &- &- &- & 16 & - & -  & -  \\
        & & & & \\
\multicolumn{2}{c}{L = 15~000 \lsol} & & & \\
        & & & & \\
${\dot{M}}_{\rm def} \times$ 1.2 & - & -& - & - & - & - & - & - & - & 8 \\
${\dot{M}}_{\rm def} \times$ 1.0 & 9 & -& - & - & - & - & 9 & 9 & 9  &- \\
${\dot{M}}_{\rm def}$ / 1.2     &  10 & -& - & - & - & - & 10 & {\it \bf  11} & -  &- \\
${\dot{M}}_{\rm def}$ / 1.4     &  12 & -& - & - & - & - & 12 & - & - &- \\
${\dot{M}}_{\rm def}$ / 1.6     &  14 & -& 14 & - & - & - & - & - & - &- \\
        & & & & \\
\multicolumn{2}{c}{L = 20~000 \lsol} & & & \\
        & & & & \\
${\dot{M}}_{\rm def} \times$ 1.2 & - & -& - & - & - & - & - & - & - & 8 \\
${\dot{M}}_{\rm def} \times$ 1.0 & 9 & -& - & - & - & - & 10 & 9 &-  &- \\
${\dot{M}}_{\rm def}$ / 1.2     &  - & -& - & - & - & - & - & {\bf 12} & -  &- \\
${\dot{M}}_{\rm def}$ / 1.4     &  - & -& - & - & 14& - & - & - & - &- \\
        & & & & \\
\multicolumn{2}{c}{L = 25~000 \lsol} & & & \\
        & & & & \\
${\dot{M}}_{\rm def} \times$ 1.4 & - & -& - & - & - & - & 7 & 7 & 7 & - \\
${\dot{M}}_{\rm def} \times$ 1.2 & - & 8& - & - & - & - & 8 & - & 8 & {\bf  8} \\
${\dot{M}}_{\rm def} \times$ 1.0 & 9 & -& - & - & - & - & 10 & 9 &-  &- \\
${\dot{M}}_{\rm def}$ / 1.2     &  10 & -& - & - & - & - & 10 & 11 & -  &- \\
${\dot{M}}_{\rm def}$ / 1.4     &  12 & -& - & - &- & 13 & 12 & - & - &- \\
${\dot{M}}_{\rm def}$ / 1.6     &  15 & -& 15 & - & - & - & 12 & - & - &- \\
${\dot{M}}_{\rm def}$ / 1.8     &  17 & -& - & - & - & - & - & - & - &- \\
        & & & & \\
\multicolumn{2}{c}{L = 30~000 \lsol} & & & \\
        & & & & \\
${\dot{M}}_{\rm def} \times$ 1.2 & - & 8 & - & - & - & - & - & - & - & - \\
${\dot{M}}_{\rm def} \times$ 1.0 & 10 & -& - & - & - & - & 10 & {\bf 10} &-  &- \\
${\dot{M}}_{\rm def}$ / 1.2     &  10 & -& - & - & - & - & 11 & - & -  &- \\
${\dot{M}}_{\rm def}$ / 1.4     &  11 & -& - & 12 &- & - & - & - & - &- \\
\hline
\end{tabular}
\end{flushleft}

\noindent
For every luminosity the combinations of models that have been
calculated are indicated. The corresponding absolute dust opacity follows
from Eq. (3). The default mass loss rate and the distance are
different for each luminosity; see Table~2. The entry in the table
is the best fitting CO abundance in units of 10$^{-4}$. Bold-faced 
numbers indicate best-fitting models discussed in Sect. 5.

\end{table*}

The shape of the calculated SED is determined completely by the input
spectrum of the underlying star, the inner dust radius (or
equivalently, the temperature of the dust at the inner radius, $T_{\rm
c}$), and the optical depth.  The optical depth is given by:
\begin{displaymath}
{\tau}_{\lambda} = \int_{r_{\rm inner}}^{r_{\rm outer}} \pi a^2 Q_{\lambda} 
\; n_{\rm d}(r) \; dr 
\end{displaymath}
\begin{equation}
 \hspace{0.5cm} \sim \frac{\dot{M} \; \Psi \; 
Q_{\lambda}/a}{R_{\star} \; v_{\infty}\; {\rho}_{\rm d} \; r_{\rm c}} \; 
\int_{1}^{r_{\rm max}} \frac{R(x)}{x^2 w(x)}\; dx
\end{equation}
where $x = r/r_{\rm c}$, $\dot{M}(r) = \dot{M} \; R(x)$ and $v(r) =
v_{\infty} \; w(x)$.  The normalized mass loss rate profile $R(x)$ and
the normalized velocity law $w(x)$ should obey $R(1)$ = 1 and
$w(\infty)$ = 1, respectively. In the case of a constant mass loss
rate and a constant velocity, the integral in Eq. (3) is essentially
unity since $r_{\rm max}$ is always much larger than 1.  The symbols
and units in Eq. (3) are: the (present-day) mass loss rate $\dot{M}$
in \msolyr, $\Psi$ the dust-to-gas mass ratio (assumed constant with
radius), $Q_{\lambda}$ the extinction coefficient, $a$ the grain size
in cm (the model assumes a single grain size), $R_{\star}$ the stellar
radius in solar radii, $v_{\infty}$ the terminal velocity of the dust
in \ks, ${\rho}_{\rm d}$ the dust grain specific density in g
cm$^{-3}$, $r_{\rm c}$ the inner dust radius in units of stellar radii
and $r_{\rm max}$ the outer radius in units of $r_{\rm c}$.  The mass
loss rates derived in G97 are based on an assumed dust velocity of
17.5 \ks, an absolute opacity at 60 $\mu$m of 68 cm$^2$g$^{-1}$ and a
dust-to-gas ratio of 0.005.

The relation between the wavelength dependent absorption coefficient
$Q_{\lambda}$ in Eq. (3) and the effective absorption coefficient $Q$ in
Eq. (1) is:
\begin{equation}
Q = \frac{\int F_{\lambda} \; Q_{\lambda} \; d{\lambda}}{\int F_{\lambda} \;
d{\lambda}}
\end{equation}
where $F_{\lambda}$ is the emerging flux from the central star and the
dust shell at infinity.  In principle $Q$ depends on the radius since
$F_{\lambda}$ is continuously changed by the dust emission but this is
only important near the inner radius where the dust temperature
changes rapidly (see Habing et al. 1994).  The molecular transitions
under discussion here originate from several tens to hundreds of
stellar radii where $F_{\lambda}$ is essentially independent of the
radial distance.

Radiative pumping of molecules is provided by thermal emission from
hot dust close to the star. In molecular emission models this is
usually represented by a blackbody of temperature $T_{\rm BB}$ and
radius $R_{\rm BB}$.  These quantities can be estimated from the
DRT-models. At each gridpoint in the DRT-model the blackbody
temperature of the radiation field is determined. In this way a
realistic estimate of $T_{\rm BB}$ and $R_{\rm BB}$ is obtained.

In G97 we showed that the near-infrared visibility curves are very
sensitive to the grain size and that a main grain size of 0.16
$\pm$ 0.01 $\mu$m could fit the visibility curves. Therefore a
grain size of 0.16 $\mu$m is adopted in all calculations.

Finally, DRT-models provide ${\tau}_{0.1}$, the optical depth at 1000
\AA\ which is needed to calculate the photoelectric heating (see G94
for details) and the dust temperature profile which is needed in the
heating rate due to the gas-dust temperature difference (see G94).

\subsection{The outer radius and photodissociation}

The outer radius of the CO shell is determined by photodissociation.
Mamon et al. (1988) have investigated this effect and found that the
CO abundance as function of radius, relative to the value close to the
star, can be approximated as:
\begin{equation}
X_{CO} = e^{-ln (2) \; (r/r_{1/2})^{\alpha}}
\end{equation}
where $r_{1/2}$ and $\alpha$ depend on $\dot{\rm M}$ and $v$ and are
tabulated by Mamon et al. (1988). Here we instead use the
analytical fit to these results by Stanek et al. (1995):
\begin{displaymath}
r_{1/2} = 5.4 \times 10^{16} \,\, \left(\frac{\dot{M}}{10^{-6}}\right)^{0.65}
\;\;\left(\frac{v}{15}\right)^{-0.55} \,\, 
\left(\frac{f_{\rm CO}}{8 \times 10^{-4}}\right)^{0.55}
\end{displaymath}
\begin{equation}
\hspace{1.0cm} + \, 7.5 \;\times 10^{15} \left(\frac{v}{15}\right) \;\; {\rm cm}
\end{equation}
and Kwan \& Webster (1993)
\begin{equation}
\alpha = 2.79 \; \left(\frac{\dot{M}}{10^{-5}} \frac{15}{v}\right)^{0.09}
\end{equation}
where $f_{\rm CO}$ is the number ratio of CO to H$_2$ molecules close
to the star, \mdot\ is the mass loss rate in \msolyr\ and $v$ the gas
velocity in \ks. The photodissociation radius of $^{13}$CO is to first
approximation assumed to be equal to that of $^{12}$CO (see Mamon et
al. 1988). The outer radius {of the CO shell is set at the radius
where the CO abundance has dropped to 1\% in the calculations}.
Equations. (6) and (7) are used with a gas velocity of 14.5 \ks, which is
the observed terminal velocity of the gas.

\section{Best guesses for the dust-to-gas ratio and CO abundance}

Two important free parameters are the dust-to-gas ratio and the
abundance of CO molecules relative to H$_2$. Both quantities can not
take arbitrary values.

There is a physical upper limit to the dust-to-gas ratio based on the
number of atoms that can condense into dust. Using the continuity
equation for the gas and the dust and assuming that the dust is 100\%
carbonaceous one finds that the theoretical dust-to-gas ratio is given
by:

\begin{equation}
\Psi = f_{\rm c} \, ({\rm C/O} - 1) \, \frac{n_{\rm O}}{n_{\rm H}} \,
\frac{12}{1.4} \, \frac{v_{\rm gas} + v_{\rm dr}}{v_{\rm gas}}
\end{equation}
where $f_{\rm c}$ is the degree of condensation of the dust and C/O is
the number ratio of carbon to oxygen atoms in the gas phase. Taking
the observed gas velocity of 14.5 \ks, and assuming a drift velocity
of 3.0 \ks\ and a cosmic oxygen abundance of 8.70 (on a scale where H
= 12.0) then the theoretically predicted dust-to-gas ratio is $\Psi$ =
 5.18 $\times$ 10$^{-3}$ $f_{\rm c}$ (C/O $- 1)$.

For disk planetary nebulae it found that C/O $<$ 4 (see
e.g. Groenewegen \& de Jong 1994 and references therein) and because
$f_{\rm c} <$ 1 by definition, it follows that $\Psi$ must be smaller
than 0.016. This upper limit is valid for all carbon
stars. Calculations show that $f_{\rm c}$ is about 0.4 (within a
factor of 2) (see Fleischer et al. 1995, Winters et al. 1994a,b) and
for IRC~+10~216 a C/O ratio of 2-3 seems appropriate (G97).  In this
case $\Psi$ is in the range 0.001 - 0.008.  In the model calculations
we use dust-to-gas ratios between 0.001 and 0.005, in agreement with
the theoretically expected range.

For $f_{\rm CO}$, defined as the n(CO)/n(H$_2$) ratio close to the
star, one may assume that in a carbon star all oxygen atoms are locked
up in CO. Using an oxygen abundance of 8.70 $\pm$ 0.1 it immediately
follows that $f_{\rm CO}$ = (10 $\pm\, 2)\, \times$ 10$^{-4}$. All the
best fitting models have derived CO abundances in this range. Note
that some of the other models discussed in the literature (Kwan \&
Linke 1982, Crosas \& Menten 1997) find a CO abundance of 6 $\times$
10$^{-4}$, which is in disagreement with the theoretically expected value.


\begin{table*}
\caption[]{Fit to secondary constraints}
\begin{flushleft}
\begin{tabular}{llllllll} \hline
Transition  & Telescope & Observed Peak & \multicolumn{5}{c}{Model Peak
Temperature(K) }\\ 
            &       & Temp. (K)  & $L$ = 10~000 \lsol & 15~000 &
20~000  & 25~000  & 30~000  \\ \hline
$^{12}$CO (1-0) & BTL &  5.1 &  4.9 &  5.5 &  5.9 &  5.5 &  5.8 \\
               & NRAO &  9.6 &  9.3 & 10.3 & 10.9 & 10.3 & 10.8 \\
               & SEST & 10.7 & 10.2 & 11.2 & 11.9 & 11.2 & 11.7 \\
               & OSO  & 10.0 & 13.2 & 14.7 & 14.9 & 14.2 & 14.5 \\
$^{12}$CO(2-1) & NRAO & 21.5 & 15.7 & 16.8 & 17.5 & 16.6 & 17.2 \\
               & SEST & 25.4 & 21.1 & 22.1 & 23.0 & 21.9 & 22.7 \\
$^{12}$CO(3-2) & CSO  & 32.5 & --   & 22.6 & -- & -- & -- \\
$^{12}$CO(7-6) & UH2.2 & 9.0 & 2.87 & 2.66 & 2.78 & 2.65 & 2.75 \\
$^{13}$CO(1-0) & BTL &  0.38 & 0.27 & 0.29 & 0.32 & 0.30 & 0.32 \\
$^{13}$CO(3-2) & CSO  & 2.9$^{(a)}$ & --   & 2.62 & -- & -- & -- \\
\hline
\end{tabular}
\end{flushleft}

\noindent
$^{(a)}$ Wang et al. (1994) give a value of 3.9 K, but this apparently
refers to the peak temperature of the horn, in this double-peaked
profile. The observed temperature at the line center is read of their
Fig. 2 and then appropriately scaled to give the main-beam temperature.

\end{table*}

\section{The CO data}

A wealth of CO data exist for IRC~+10~216. In Table~1 we list the data
that are used as constraints here. Except for the $^{12}$CO and
$^{13}$CO J = 6-5 spectra, all data are taken from the literature.

The $^{12}$CO and $^{13}$CO J = 6-5 observations were carried out
using the James Clerk Maxwell Telescope (JCMT), located on Mauna Kea,
Hawaii.  The detector was an SIS receiver (RxG) available to JCMT via
a collaborative agreement with the Max-Planck-Institut f\"ur
Extraterrestische Physik in Garching (Germany).  A gunn
oscillator yields continuous coverage across the 650-692 GHz band,
which includes the $^{12}$CO and $^{13}$CO J = 6-5 lines.  The backend
was an acousto-optic spectrometer.  Before the observations
were made, the planets Jupiter and Mars were used to determine the
beamshape during good weather with a 30\% zenith transmission.  It was
found that, although the JCMT beam is complex at these high
frequencies, it can be well described by a two-component beam with
55\% of the power in the 7\arcsec\ FWHM diffraction component and 45\%
of the power in a 18\arcsec\ FWHM beam.  The overall beam efficiency
in this compact beam is 27\% and was determined by taking spectra of
Jupiter and Mars.  When taking all errors into account we expect the
calibration error to be less than 20\%.

The $^{12}$CO J = 6-5 observations were made on April 21 and 25, 1995.
The pointing was done using the continuum receiver UKT14.  The
relative pointing difference between the RxG and UKT14 receivers was
determined using very bright sources and was accurately known.  After
the telescope was successfully pointed at the source using UKT14, a
switch was made to RxG and the relative pointing correction was
applied.  During the observations IRC~+10~216 was always above 70
degrees elevation.  The zenith sky transmission was 17\% on April 21
and 19\% on April 25 yielding system temperatures of 17~000 and
22~000~K respectively. Four scans of 4 min each were made, three of
which were on April 21, for a total integration time of 16 min. On April
21 the on-source spectra were part of a 5-point map with a beam
separation of 8\arcsec; on April 25 the on-source spectrum was part of
a 5-point map with a 10\arcsec\ beam separation.

In Figs. 1-5 one can notice a bump in the spectrum centered near a
velocity of 0 \ks. It is not present in the J = 6-5 spectrum of Crosas
\& Menten (1997). We checked the JPL line catalog and no suitable
candidate could be found. We also checked for lines coming in from the
sideband with the same result. In any case this unidentified line does
not influence any conclusions regarding the (fitting of the) J = 6-5
profile.

The $^{13}$CO J = 6-5 observations were made on April 30, 1995.  The
pointing was done as described above for the $^{12}$CO observations.
During the observations IRC~+10~216 was between 65 and 70 degrees
elevation.  The zenith sky transmission was 32\% yielding a system
temperature of 10~000~K.  Seven scans of 8 min each were made for a
total integration time of 56 min.  No mapping was done in this line.

The above data were added after visual inspection of the individual
scans and base line subtraction.  The rms noise in the final $^{12}$CO
J = 6-5 spectrum is 2.4 K while the line is about 45~K. The rms noise in
the final $^{13}$CO J = 6-5 spectrum is 0.8 K while the line is about 7~K.

The primary constraints in this paper are the J = 6-5 profiles
discussed above, the $^{12}$CO J = 4-3 profiles from Williams \& White
(1992) and $^{12}$CO and $^{13}$CO J = 1-0, 2-1 and 3-2 profiles
observed at IRAM and JCMT and discussed in Groenewegen et
al. (1996). In addition a selection of other data was taken from the
literature in particular that taken with smaller telescopes.

\begin{figure}
\centerline{\psfig{figure=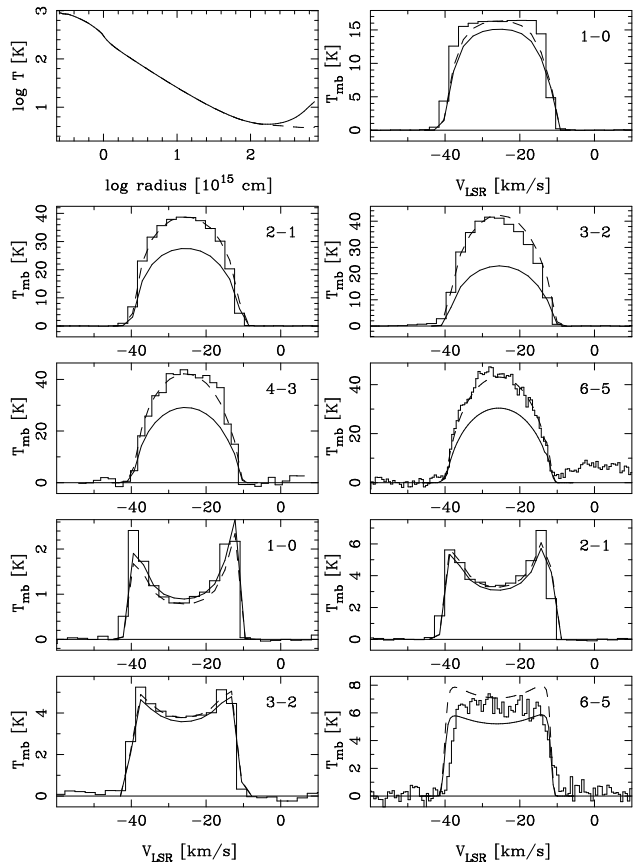,width=8.4cm,angle=-90}}
\vspace{-3mm}
\caption[]{Model with $L$ = 10~000 \lsol, and default parameters for
\mdot, $\Psi$ and $Q$.  Upper left panel indicates gas temperature
(solid line) and excitation temperature of the $^{12}$CO (1-0)
transition (dashed line). The rise in the gas temperature at large
distances from the star is due to photoelectric heating. 
Other panels indicate $^{12}$CO (upper part)
and $^{13}$CO (lower four panels) line profiles. In each case the
histogram indicates the observations. They are the ``primary
constraints'' of Table~1. The solid line indicates the model, while the
dashed line represents the model scaled to the observed maximum, in
order to better compare the line shapes.  
}
\end{figure}




\begin{figure}
\centerline{\psfig{figure=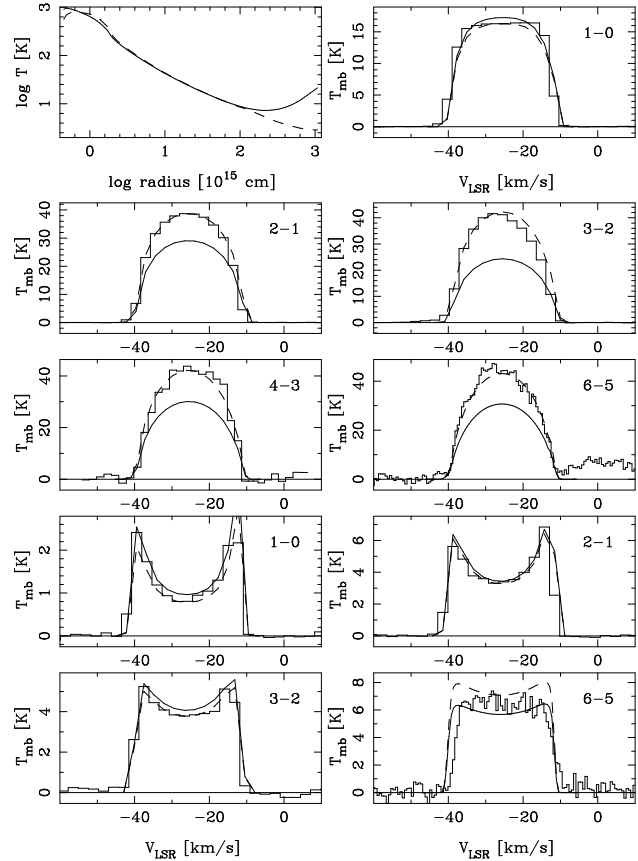,width=8.4cm,angle=-90}}
\caption[]{Model with $L$ = 30~000 \lsol, and default parameters for
\mdot, $\Psi$ and $Q$.
}
\vspace{-3mm}
\end{figure}

\begin{figure}
\centerline{\psfig{figure=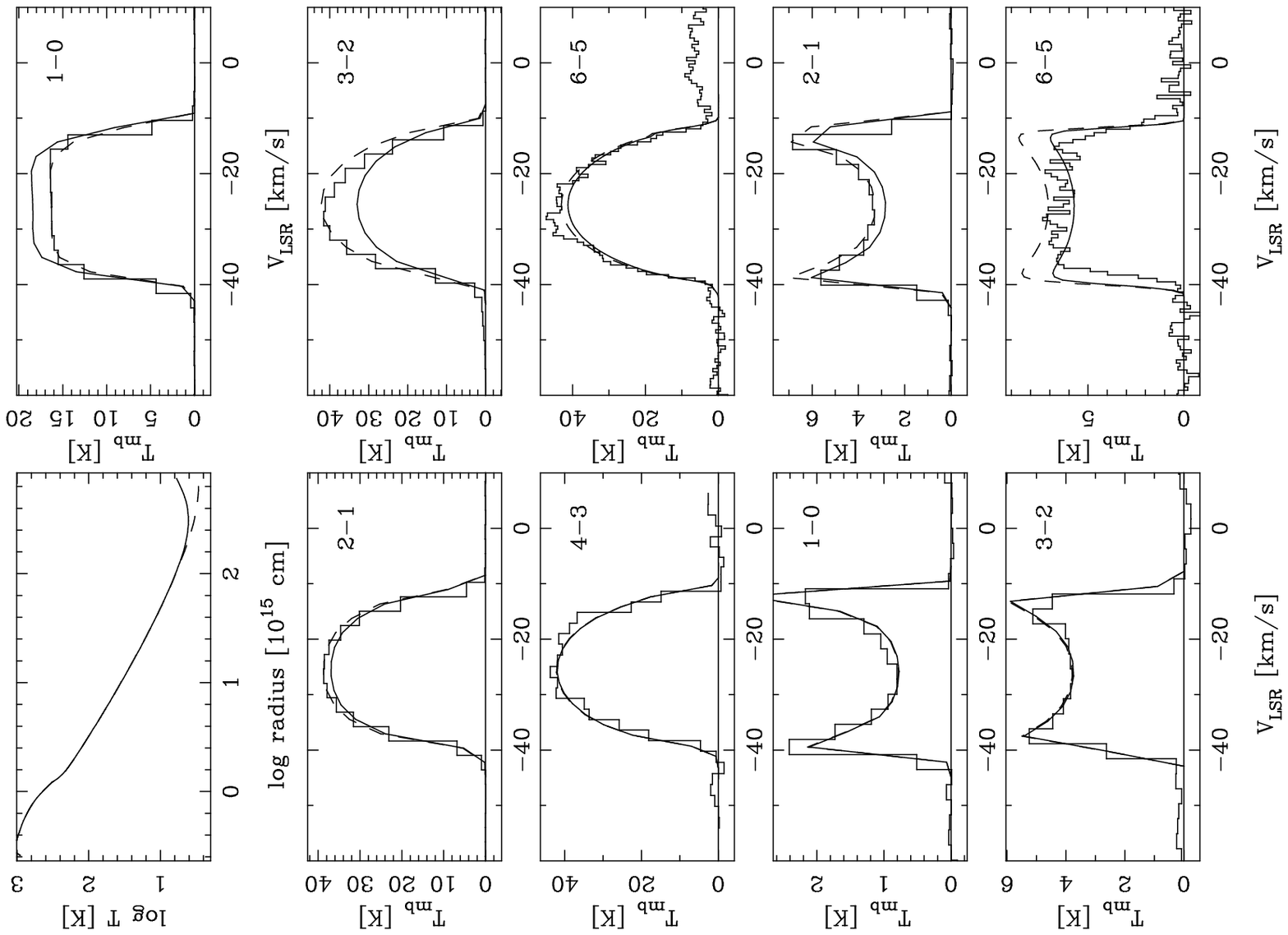,width=8.4cm,angle=-90}}
\caption[]{Best fitting model with $L$ = 10~000 \lsol.
}
\end{figure}

\begin{figure}
\centerline{\psfig{figure=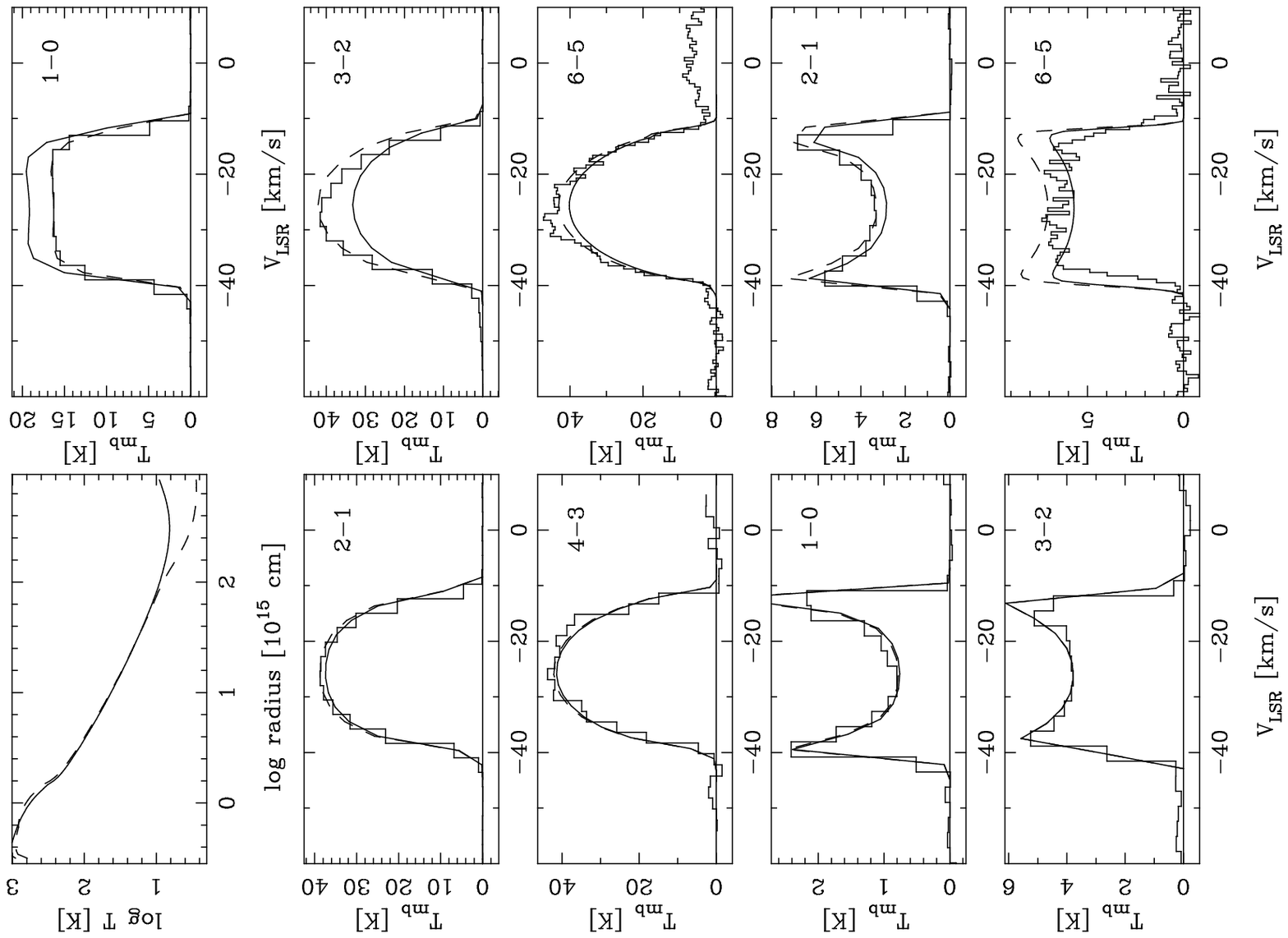,width=8.4cm,angle=-90}}
\caption[]{Best fitting model with $L$ = 15~000 \lsol.
}
\end{figure}

\begin{figure}
\centerline{\psfig{figure=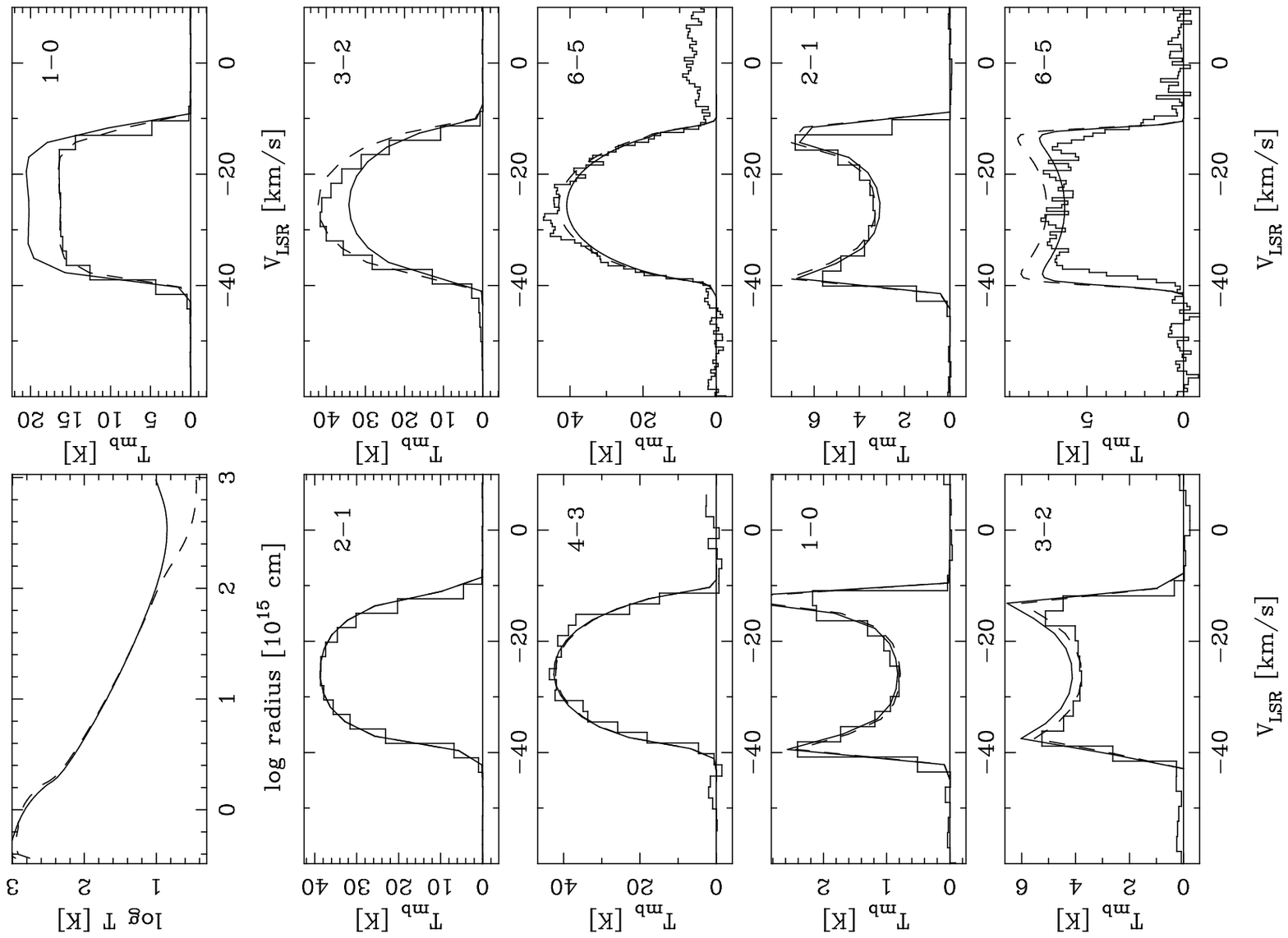,width=8.4cm,angle=-90}}
\caption[]{Best fitting model with $L$ = 20~000 \lsol.
}
\vspace{-3mm}
\end{figure}



\section{Model calculations}

A large grid of models was calculated, varying the luminosity (and the
distance correspondingly), the mass loss rate, the dust-to-gas ratio,
the absolute dust opacity and f$_{\rm CO}$ while fixing f$_{\rm
CO}$/f$_{\rm ^{13}CO}$ at a value of 44, a value accurately known from
optically thin molecular lines (see Kahane et al. 1992).  The velocity
law of the gas is parameterised by $v(r)$ = 14.5 \ks\ (1 -- 2.73
$\times$ 10$^{14}$ cm ($d$/135 pc)$/r)^{0.5}$.


From the dust modeling in G97 the following parameters were derived:
$Q$ = 0.0419; $a$ = 0.16 $\mu$m, inner (dust) radius = 3.17 $\times$
10$^{14}$ cm (d/135 pc); $T_{\rm BB}$ = 860 K; $R_{\rm BB}$ = 6.18
$\times$ 10$^{14}$ cm (d/135 pc); ${\tau}_{0.1}$ = 1.84 $\times$
10$^{16}$ cm (d/135 pc)$/r$, with $r$ the radial distance to the star.
Given the observed integrated flux, Table~2 lists the distance
corresponding to several chosen values of the luminosity, and the mass
loss rate following from the dust model.

The actual combinations of parameters used in the various CO model
calculations are listed in Table~3. For example, take the entry with
$L$ = 20~000 \lsol, ``${\dot{M}}_{\rm def}/1.4$'' and
``$\Psi/1.67$''. The default mass loss rate for that luminosity is
2.54 $\times$ 10$^{-5}$ \msolyr, and the default dust-to-gas ratio is
0.005 (see Table~2).  This therefore means that the model was run with
\mdot = 1.81 $\times$ 10$^{-5}$ \msolyr\ and $\Psi$ = 0.0030. From
Eq. (3) it follows that the dust opacity was (1.4 $\times$
1.67) $\times$ 68 = 160 cm$^2$gr$^{-1}$. For every model combination
the CO abundance ratio was tuned to give the best possible fit to both
the $^{12}$CO and $^{13}$CO line profiles marked as ``primary
constraint'' in Table~1. The $^{13}$CO abundance was fixed at $(f_{\rm
CO}/44)$ as explained above.

Figures 1-2 present the fits to the ``primary constraints'' for models
with the default values for the mass loss rate, opacity and
dust-to-gas ratio for the two extreme luminosities. Larger luminosities result
in higher predicted brightness temperatures. This is most clearly seen
in the $^{12}$CO(1-0) profile. Overall however, the effect of
luminosity appears not be very significant, and therefore only the
extreme models with $L$ = 10~000 and 30~000 are shown. 
It is also clear that these models (and the models with intermediate
luminosities) fail to fit the ``primary constraints''.

Figures 3-5 present the best fitting models for three different
luminosities. These models have a bold-faced entry in Table~3. For
every luminosity a model can be found that fits the ``primary
constraints'' almost equally well. The high-luminosity models are not
shown because they fit qualitatively equally good, and such high
luminosities can be ruled out on other grounds (see later).  In all
cases the models predict brightness temperature for the $^{12}$CO
(3-2) line that is too low.  This may indicate a systematic
uncertainty in the calibration (also see discussion in Sect. 5.1.1).
Also, all the models predict a $^{12}$CO J = 1-0 line that is too
bright. This difference is lowest for the 10~000 \lsol\ model, where
the difference is well within the calibration uncertainty. The
remaining discrepancy is the $^{13}$CO (6-5) line. The absolute
intensity is well reproduced but the line shape does not fit very
well.  None of the models calculated predict a flat-topped $^{13}$CO
(6-5) line. Also the recent model calculation of Crosas \& Menten
(1997) predicts a slightly double-peaked profile. Note that their
model does not reproduce the absolute intensity neither.

It was verified that this discrepancy is not due to a pointing
off-set. A model calculated 3.5\arcsec\  off-centre
still produced a slight double-peaked profile. A model at 7\arcsec\  off-set
produced at flat-topped profile but the absolute intensity was a
factor of 3 below the observed one.

This discrepancy hints to the fact that possibly one of the
fundamental assumptions in the model may no longer be full-filled:
spherical symmetry or homogeneity. This can in principle be addressed
by high spatial resolution, high frequency mapping observations. 
Problems associated with these high frequency observations are that
the beam is usually small, and random pointing errors can already be a
significant fraction of the primary beam, and the influence of the error
beam, which makes interpretation not straightforward.

\subsection{The other constraints}

\subsubsection{On-source data}

We compared the best fitting model at each luminosity with the line
intensities and line profiles for all on-source observations that were
not used as primary constraints (see Table~1). The peak intensities
predicted by the model are listed in Table~4, together with the
observational data from the respective papers.

The reasoning behind this exercise is that these other J = 1-0 and 2-1
observations were performed with smaller telescopes with large beam
sizes, and so these observations sample a greater extent of the envelope
than the higher resolution IRAM observations.

The J = 1-0 BTL, NRAO and SEST data are equally well fitted by the five
models, given the typical 1$\sigma$ uncertainties of 10\% in the
observations. In fact, all three observations would point to a
luminosity between 10~000 and 15~000 \lsol. Regarding the OSO
observations there must have been a calibration error. Indeed,
Huggins et al. mention that no beam efficiency correction is applied
and that the observations have been scaled to an earlier 1982 OSO
observation.  A more recent OSO observation (Olofsson et al. 1993)
indicates a peak temperature of 12.5 K which is low but formally in
agreement with the models. The $^{13}$CO J = 1-0 BTL observation agrees
reasonably well with the models.

The J = 2-1 NRAO observations are barely in agreement with the
models. Other NRAO observations find similarly high on-source
temperatures (Wannier \& Sahai 1986, Wannier et al. 1990). The SEST
data agrees with the models.

With respect to the J = 7-6 observations there is a discrepancy
of more than a factor of three. We have no explanation for this. 
This is the only existing 7-6 observation of this star, and we suggest
that this transition be re-observed. As the J = 6-5 data is well
fitted, there is no reason to believe that our model is in error by
such a large factor.

The referee kindly pointed us to the paper of Wang et al. (1994) which was
not considered originally and discusses $^{12}$CO and $^{13}$CO (3-2)
data taken with the CSO. For one model, the relevant temperatures
were calculated, and are included in Table~4. The $^{13}$CO (3-2) model
prediction is in excellent agreement with the observation, while the 
$^{12}$CO (3-2) model prediction is significantly lower.

Williams \& White (1992) also present $^{12}$CO and $^{13}$CO (3-2)
data taken with the JCMT, and can therefore be directly compared to our
observations. For the $^{12}$CO line they find a main-beam
temperature of 32 K at 2\arcsec\ off-set, and estimated an on-source
temperature of about 37 K. We find 41 K, in excellent agreement. For
$^{13}$CO they find a line-center temperature of 3 K (at a position
$\pm$5\arcsec), while we found 3.8 K. This agreement is acceptable,
and it should be pointed out that their spectrum has a considerable
lower S/N than ours.

In conclusion, the two sets of JCMT data are in agreement with each
other. For both CSO and JCMT the predicted $^{12}$CO (3-2) line is
less intense than observed. The same is found by Crosas \& Menten
(1997) who compare their model to the Wang et al. data and find a
model temperature in agreement with our value in Table~4 and hence
lower than observed.  In Sect. 5.0 it was suggested that this might be
due to a calibration problem.  This would imply calibration problems
at both CSO and JCMT, which appears unlikely. On the other hand a
physical cause also appears unlikely. As temperature and density are
smooth functions of radius, the profiles are also expected to be
smooth function of increasing J-level. Since the $^{12}$CO J=2-1 and
4-3 levels are well fitted, there is no a-priori reason to believe that
the 3-2 level would be off. One could assume that at a radius, where
the emission from the 3-2 level is strongest, the density is lower
than predicted from the standard model, but this would also then imply
a different $^{12}$CO/$^{13}$CO ratio at that particular radius only
as the the $^{13}$CO 3-2 profile is predicted correctly.

\begin{figure*}
\centerline{\psfig{figure=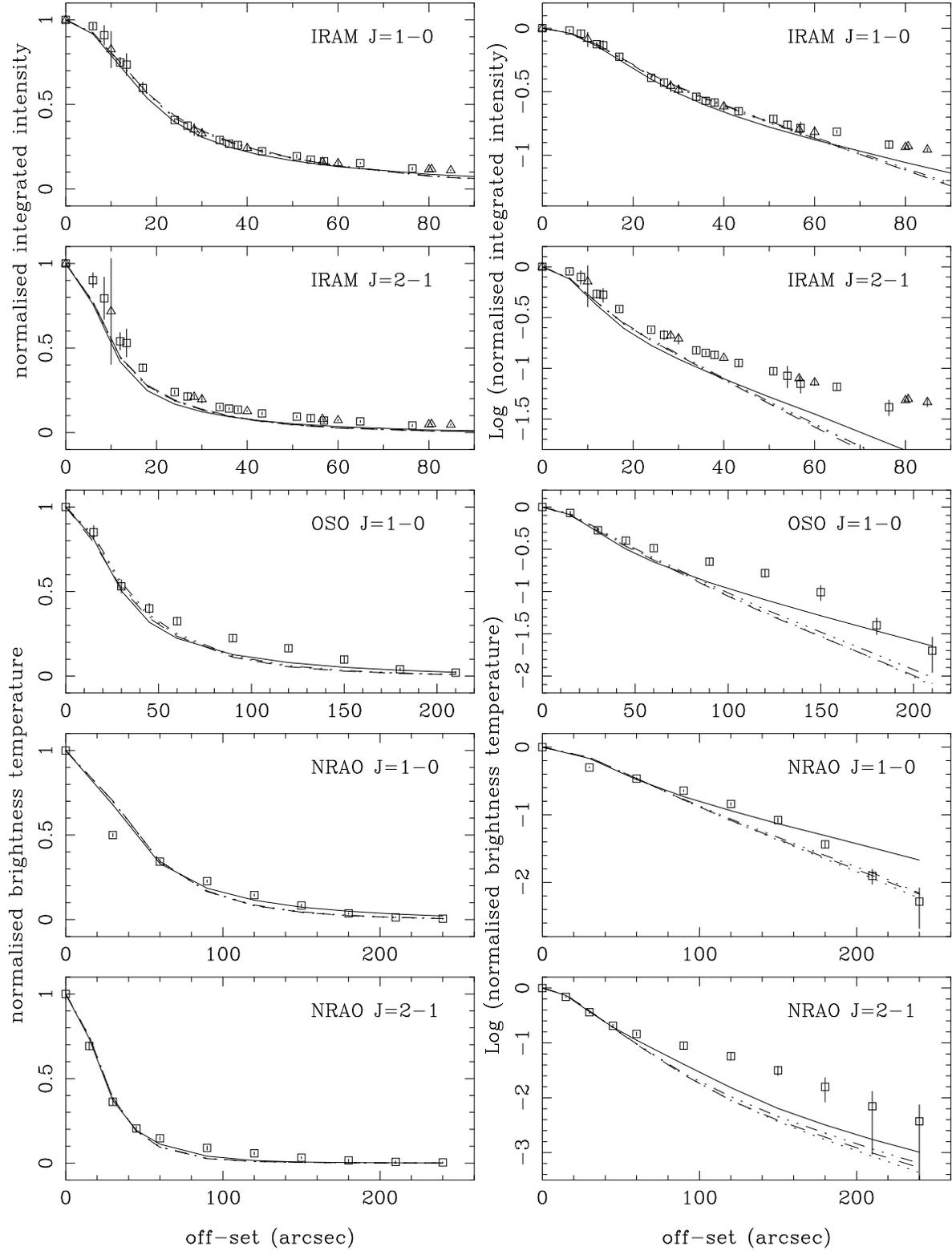,width=15.8cm,angle=0}}
\vspace{-2mm}
\caption[]{Comparison between observations (symbols with error bars)
and models (lines) as a function of radial offset. IRAM observations
come from Truong-Bach et al. (1991; squares), and Groenewegen \&
Ludwig (1998; triangles). The OSO and NRAO observations are from
Huggins et al. (1988). Best fitting models are for $L$ = 10~000
(solid), 15~000 (dashed), 25~000 (dotted), 30~000 \lsol\
(dash-dot-dot-dotted), and default values for photoelectric heating
and constant mass loss rate. For the IRAM observations normalised
integrated intensities, and for the OSO/NRAO observations normalised
peak brightness temperatures are shown, on both a linear (left hand) 
and logarithmic scale (right hand).
}
\end{figure*}

\subsubsection{Mapping data}

In Fig.~6 the four best fitting models are compared to the mapping
data of Huggins et al. (1988), Truong-Bach et al. (1991) and
Groenewegen \& Ludwig (1998), both on a linear and logarithmic scale.
For the IRAM observations integrated intensities are compared, while
for the OSO and NRAO observations by Huggins et al. only peak
intensities are given in their paper. There exists at map in the
$^{12}$CO J = 3-2 line (Williams \& White 1992), but they find that
the emission appears not to be extended with respect to the beam out
to their maximum off-set of 50\arcsec. Therefore we did not consider
this data.

A first remark is that there is extremely good agreement between the two
sets of IRAM observations. The error bars plotted indicate the
variation in the ratio of the integrated intensity relative to the
on-source value at the different map positions for a given radial offset.

At first glance there appears to be good agreement between the models
and observations when plotted on a linear scale, but, as was noted by
Huggins et al. (1988), clear discrepancies are seen when plotted on a
logarithmic scale. Although the absolute calibration uncertainties are
of order 10\% (and larger in the case of the OSO J = 1-0 observation),
the relative intensities are much better determined and only limited by noise.

The models and observations start to deviate at approximately
50\arcsec\ offset from the center position, the exact value depending
on the beam-size. There are two possible explanations for this. First,
the mass loss rate may have been higher in the past. Second, the
photoelectric effect may be more important than assumed.

Models with an increased mass loss in the past have been constructed
to fit the data for luminosities of 15~000 and 30~000 \lsol. A good
fit is obtained when the mass loss rate is higher by a factor of 5 for
radial offsets \more 50\arcsec. These models are shown in Fig.~7 on
a logarithmic scale only. 
The fit to the IRAM data is very good. That to the NRAO J = 2-1 
and OSO J = 1-0 data is acceptable. On the other hand there is an
obvious discrepancy between the OSO and NRAO J = 1-0 data. In the
present model we assumed that the increase in the mass loss rate is
constant and extends to the outer radius. Given the presently
available relatively poor data for radial offsets \more 90\arcsec,
and the discrepancies for radial offsets \more 150\arcsec, it can not
be excluded that the mass loss rate is lower again in the outer parts
of the envelope. 
The error in the radius of the on-set of the enhanced mass loss rate
and the factor by which this is so are approximately 10\arcsec\ and
less than a factor of 2, respectively. From the radial dependence
observed with IRAM compared to the constant mass loss rate case it is
immediately clear that there is a break in the observations between 40
and 60\arcsec.  It was verified numerically that a model with an
increase in mass loss rate of a factor of 3 did not fit the data.
Although we do not claim that our best-fitting model is unique, this
gives some constraints on the associated uncertainties of this model.

As the increase in the mass loss rate occurs at offsets much larger
than the respective beam sizes, negligible changes are expected in the
on-source fluxes. This was verified numerically.

Models with increased photoelectric heating were also considered (for
the case of a constant mass loss rate and $L$ = 15~000 \lsol). Even a
model with the parameter $(G \times Y$; see Sect.~2) a factor 
six larger than the
default value could not fit the IRAM data, although it could fit some
of the NRAO/OSO data. The reason is, that for the photoelectric effect
to be important even at 50\arcsec\ from the star the interstellar
radiation field must be very strong.  Apart from the fact that the
model with increased photoelectric effect does not fit the IRAM data,
such a model appears also unphysical, as for a distance of 150 pc and 
its galactic latitude IRC~+10~216 is at 110 pc from the galactic
plane, where one does not expect an increase in the diffuse
interstellar radiation field by such a large amount.

\begin{figure}
\centerline{\psfig{figure=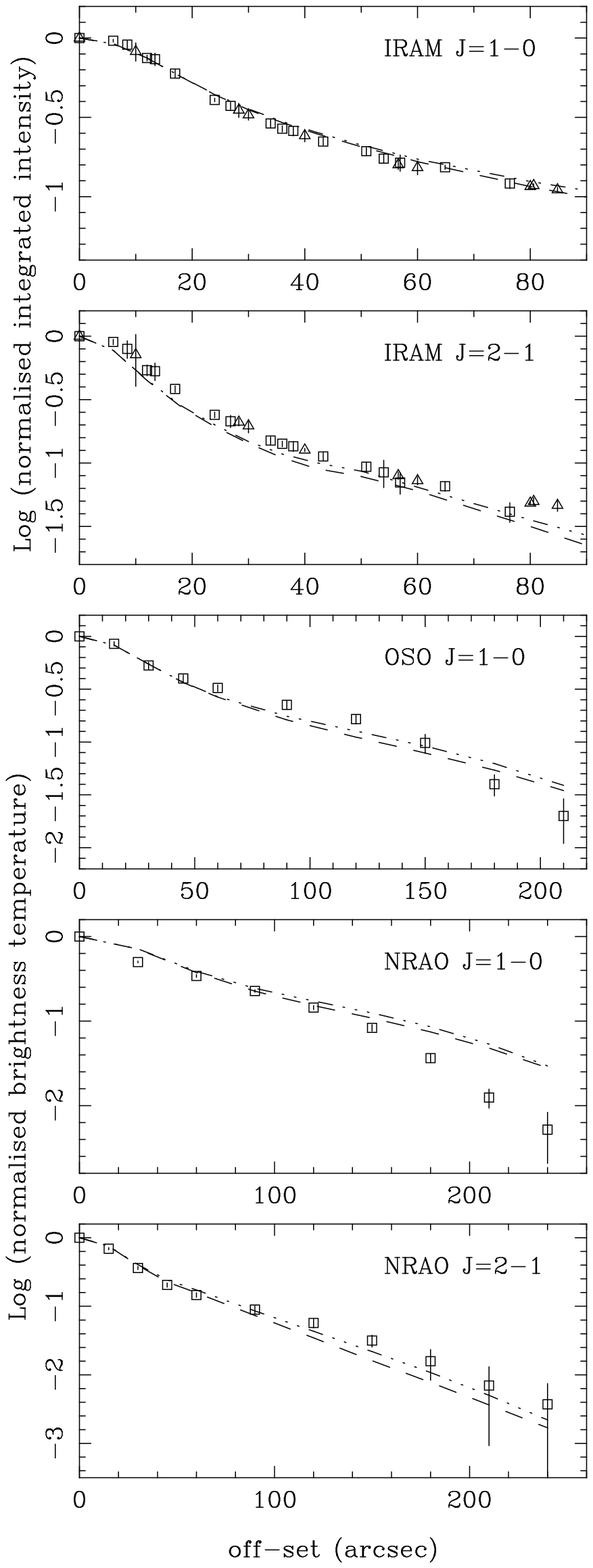,width=7.8cm,angle=0}}
\caption[]{As previous figure but now for models including a higher
mass loss rate in the past, as discussed in the text. 
$L$ = 15~000 (dashed), 30~000 \lsol\ (dash-dot-dot-dotted).
}
\end{figure}

\section{Visibility curves}

In principle, interferometric observations yield additional
information on the structure of the circumstellar envelope. In an
article in the JCMT newsletter,  Hills (1996) briefly mentions that
interferometric observations at the wavelength of the J = 4-3 line had
been performed with the JCMT-CSO interferometer. To provide  comparison
with future observations, we have calculated the so-called visibility
curves at the center wavelengths of the J = 1-0 up to 6-5 $^{12}$CO
transitions for the best fitting models with constant mass loss
rate. For a brief summary on what a visibility curve is, and how they
are evaluated, we refer to Sect. 2.1 of G97. 
The results of the calculations are plotted in Fig.~8.

The model with mass loss variations as described above yield results
indistinguishable from the constant mass loss rate case. This is due to
the fact that these mass loss variations occur at large radii.

At the same time this implies that the visibility curves may be most
sensitive to the geometry of the CO shell.  Asymmetries would be
revealed by obtaining the visibility curves for different position
angles.

\begin{figure*}
\centerline{\psfig{figure=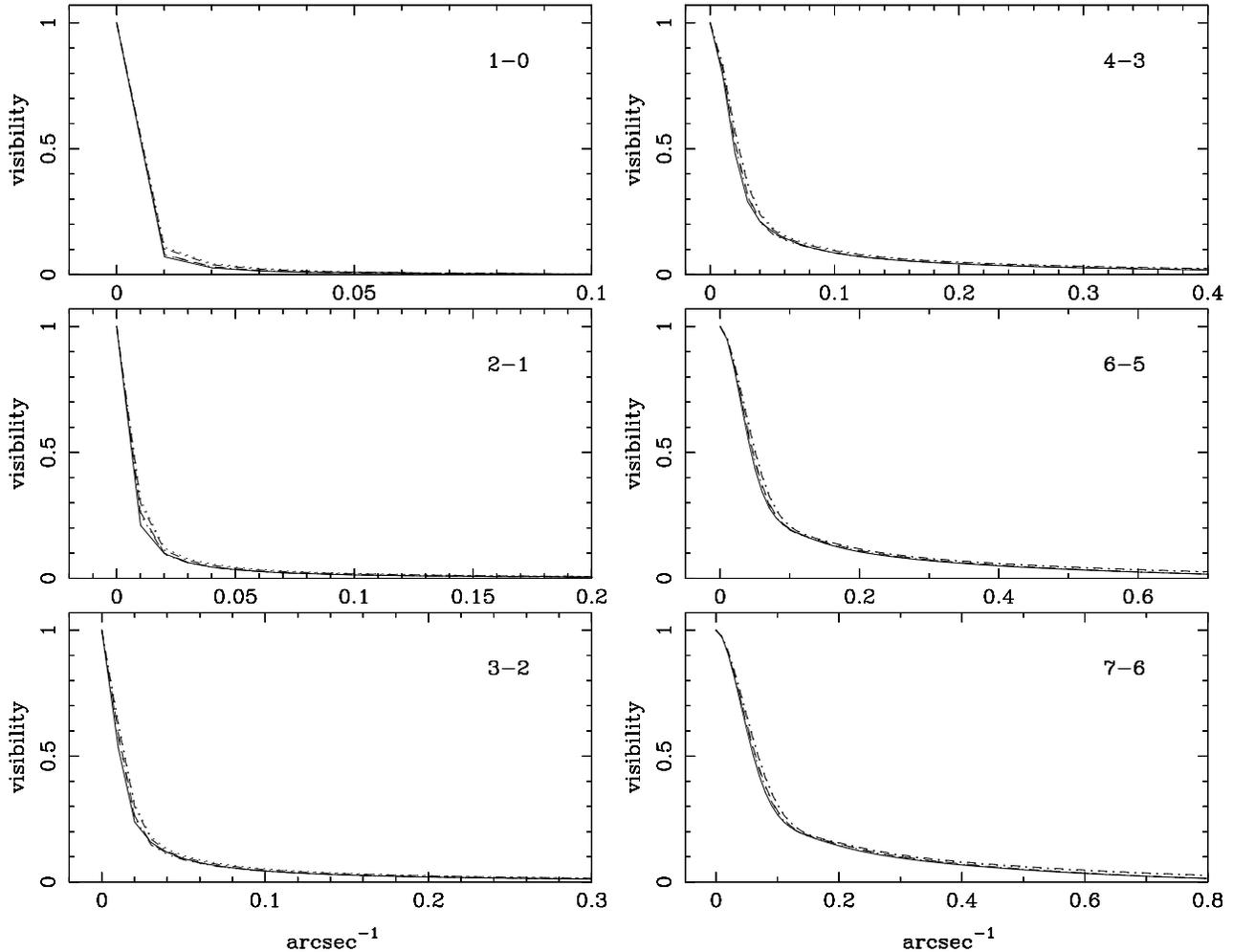,width=17.0cm,angle=-90}}
\vspace{-2mm}
\caption[]{Visibility curves for the best fitting models with constant
mass loss rate, $L$ = 10~000 (solid), 15~000 (dashed), 20~000
(dot-dashed), 25~000 (dotted), 30~000 \lsol\ (dash-dot-dot-dotted),
for different transitions. The calculations are performed with a
resolution of 0.01 arcsec$^{-1}$.
}
\end{figure*}

\begin{figure}
\centerline{\psfig{figure=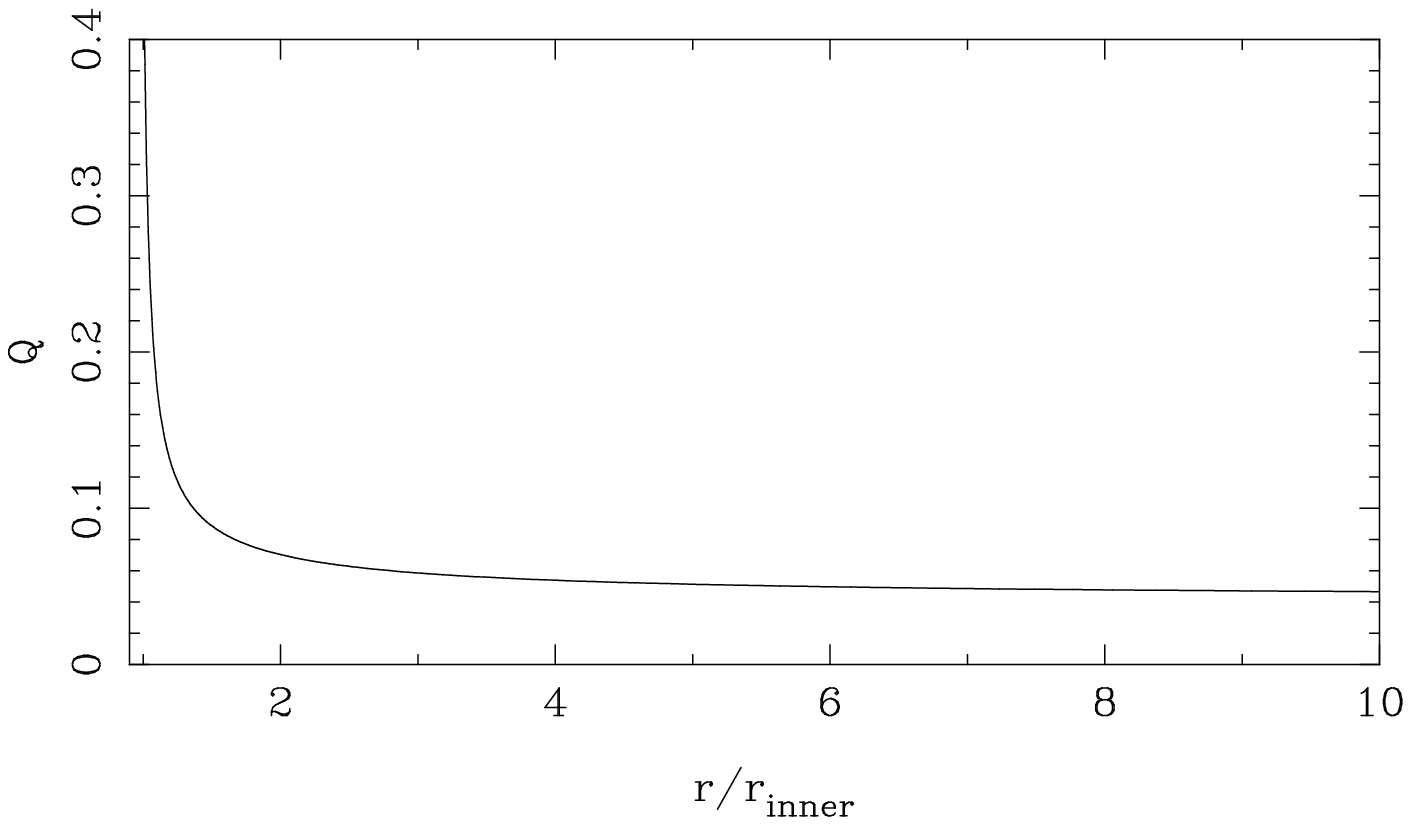,width=8.5cm,angle=0}}
\vspace{-2mm}
\caption[]{Radial dependence of the gas-dust momentum transfer
efficiency $Q$.
}
\vspace{-3mm}
\end{figure}

\section{Discussion}

\subsection{General comments}

We have made the most consistent modelling of the CO and dust envelope
around IRC~+10~216 so far. In particular we did not assume that the
gas-dust momentum transfer rate is a free parameter but constrained it
from our previous dust modelling. The physical range of the
dust-to-gas ratio and the CO abundance are considered, contrary to
some other models that find unrealistically low values for the CO
abundance.

Contrary to Sahai (1987) and Truong-Bach et al. (1991), and in
agreement with Crosas \& Menten (1997), we find no evidence for a hot
core in the inner 6\arcsec\ region. Specifically with respect to the
Truong-Bach et al. result, Crosas \& Menten suggest that they took too
few rotational levels into account and therefore underestimated the
rotational cooling in the inner part. Truong-Bach et al. inference was
based on an old J = 6-5 observation (Koepf et al. 1982). For the
best-fitting 15~000 \lsol\ model we find a peak temperature of 5.6 K
in the corresponding 35\arcsec\ beam, almost a factor of three below
the reported value.

Cernicharo et al. (1996) discuss an ISO LWS grating observation of
IRC~+10~216. On top of the dust continuum there is a forest of HCN
and CO lines. They infer that temperatures of
order 700-1500 K are needed to explain these observations. Specifically 
they adopt a temperature of 1200 K for radii smaller than 6 $\times$
10$^{14}$ cm. From our self-consistent temperature determination we
find a gas temperature at the inner radius (3.2 $\times$ 10$^{14}$
cm for a distance of 135 pc) of 1100 K. This agreement is encouraging.

In Groenewegen et al. (1997) we mapped IRC~+10~216 in the 1.3 mm
continuum and found evidence for phases of enhanced mass loss in the
past at radial distances of 5 and 20\arcsec\ from the star. Although
we find evidence for a higher mass loss rate in the past from the CO
observations, the radial scales are different. This is not necessarily
incompatible. First, the largest variation found from the continuum
map is only a factor of two at most and takes place from 10 to
30\arcsec\ at most. This may be not detectable in the CO. Secondly, as
the continuum emission is detected at a significant level only out to
45\arcsec\ it is not possibly to prove or disprove if the variations
seen in CO are also seen in the continuum emission.

\subsection{The influence of the radial dependence of $Q$}

Crosas \& Menten (1997) find that they require a higher value for $Q$
in the inner region, specifically $Q$ = 0.025 for $r$ \less 10$^{16}$
cm (that is 20 times their inner radius) and $Q$ = 0.018 for $r >
10^{16}$ cm. This touches upon the expectation (see Sect. 2.1) that
$Q$ should be a function of distance to the star because the effective
wavelength of the emerging spectrum is shifting to longer wavelength
due to the influence of dust.  
Crosas \& Menten claim that their adopted behaviour for $Q$ is in
agreement with recent calculations, e.g. Habing et al. (1994), which
showed that $Q(r)$ is a strongly decreasing function of $r$. We in
fact took the results by Habing et al. to assume a constant $Q$
throughout the envelope. We will now address the influence of this
assumption by taking the radial dependence of $Q$ into
account. Figure~9 shows the radial dependence of $Q$ as calculated
from the best-fitting dust model. It shows that $Q$ is rapidly
dropping from large values to a value of 0.0419 adopted in all
calculation so far.  However, and contrary to the argument used by
Crosas \& Menten, it is clear that $Q$ starts to deviate from its
value at infinity only for radial distances much smaller than
10$^{16}$ cm. In Table~5 on-source peak temperatures for various
$^{12}$CO transitions are collected for the best-fitting model with L
= 15~000 \lsol, and different assumptions about $Q$. The list includes
the J = 9-8 and 12-11 transitions which can not be measured from the
ground, but can be from space.  Shown are the standard case with a
constant $Q$ down to the inner radius, and various models where the
radial dependence of $Q$ is taken into account down to a certain inner
radius. For inner radii smaller than 5 $\times$ 10$^{14}$ cm the model
experiences some convergence problems.  Since a greater value for $Q$
means more heating (see Eq.~1) it is expected that models which take
into account the radial dependence of $Q$ lead to higher temperatures
in the inner region and hence more emission from the high-J
transitions.

\begin{table*}
\caption[]{Influence of the radial dependence of Q. On-source
peak temperatures for various transitions.}
\begin{flushleft}
\begin{tabular}{llrrrrr} \hline
Transition & Beam & $r_{\rm inn}$ = 3.2 $\times 10^{14}$ cm & 3.2
$\times 10^{15}$ & 1.7 $\times 10^{15}$  & 7.9 $\times 10^{14}$ & 5.0
$\times 10^{14}$ \\
           & (\arcsec) & $Q(r_{\rm inn})$ = 0.0419 & 0.0466 & 0.0503 &
0.0629 & 0.0838 \\ \hline
1-0 & 21   & $T_{\rm mb}$ = 19.2 & 18.3 & 18.9 & 19.1 & 19.2 \\
2-1 & 12.5 & 37.4 & 32.9 & 36.2 & 37.9 & 38.2 \\
3-2 & 14.3 & 33.2 & 29.4 & 32.1 & 34.0 & 34.4 \\
4-3 & 11.0 & 41.4 & 34.5 & 38.8 & 42.5 & 43.6 \\
6-5 &  7.0 & 62.0 & 45.0 & 54.2 & 63.8 & 67.3 \\
7-6 & 45   &  2.7 &  2.4 &  2.6 &  2.8 &  2.9 \\
9-8 & 80   & 0.53 & 0.43 & 0.49 & 0.56 & 0.59 \\
12-11 & 60 & 0.55 & 0.37 & 0.46 & 0.58 & 0.63 \\ \hline
\end{tabular}
\end{flushleft}
\end{table*}

The result of the calculations show two things. For the low-J
transitions, where the radial dependence of $Q$ should be of no
importance since the excitation temperatures are low, it is important
to choose the inner radius not too big as this will result in an
underestimate of the brightness temperature. This might also be part
of the explanation for the need of an inner hot core by Truong-Bach
et al. They assumed an inner radius of 2\arcsec\ or 6.0 $\times$
10$^{15}$ cm for their adopted distance. This results in an
underestimate of the true brightness temperature. If they had chosen a
smaller inner radius (equal to the inner dust radius, which is in fact
a physically plausible choice) the need, if any, for a hotter inner
core would have been less.

For the J = 3-2 and 4-3 transition the effects are  smaller than 5\% in
the present case. The J = 7-6 result shows that the effect becomes
important for higher levels, and smaller beam sizes. As the high-J
levels are generally measured with large beams, the effects are again
moderate (of order 10\%).

\subsection{The overall best fit}

Can a best model be identified? Not from the dust modelling alone as
is mentioned in G97, since the shape of the spectral energy
distribution is determined by the optical depth and the luminosity is
a scaling parameter (see Table~2). The reason to consider CO
observations as well, is that the heating rate is a non-linear
function of luminosity (see Eq.~1). It turns out that for a plausible
range of luminosities an almost equally best-fitting model to the
primary constraints can be found.

Some of the published CO J = 1-0 observations point to a luminosity
\less 15~000 \lsol. An independent constraint is the period-luminosity
relation for carbon miras (Groenewegen \& Whitelock 1996). For a
period of 649 days (Le Bertre 1992) it implies a luminosity of 9850
\lsol. The uncertainty in the period of about 10 days implies an
uncertainty of about 150 \lsol. The uncertainty in the fit itself is
0.26 in $m_{\rm bol}$ which implies an error of 30\% in luminosity.
Thus from the $PL$-relation one derives therefore a luminosity between
7~700 and 12~500 \lsol.

An additional constraint comes from the observed terminal gas
velocity. The assumption that radiation pressure on
dust drives the outflow puts constraints on the mass loss rate
and other parameters. Basically the formula $\dot{M} \; (v_{\infty}
-v_0) = {\tau}_{\rm F} \; \frac{L}{c} \; \left(1 -
\frac{1}{\Gamma}\right)$ must be obeyed (see Netzer \& Elitzur 1993,
Ivezi\'{c} \& Elitzur 1995, Groenewegen et al. 1998), with $v_0$ the
gas velocity at the inner radius, ${\tau}_{\rm F}$ the flux-weighted
optical depth (known from the DRT modelling), and $\Gamma$ is the
ratio of radiation pressure to gravitational pull which scales like
$\Gamma \sim \frac{Q \; L \; \Psi}{a \; {\rho}_{\rm d} \; M_{\star}}
\; \frac{v_{\infty}}{ v_{\infty} + v_{\rm drift} }$. Details on how
this system of equations is solved iteratively is explained in
Groenewegen et al. (1998).  The equation allows one to verify that the
mass loss rate and dust-to-gas ratio derived from the CO modelling are
consistent with those derived from the DRT modelling using the above
equation and procedure.  Unknowns are the mass of the star
($M_{\star}$ = 0.8 \msol\ adopted), the gas velocity at the inner
radius ($v_0$ = 1 \ks\ adopted). The dust opacity is the one given by
the best-fitting CO model for each luminosity. The results are listed
in Table~6.

\begin{table}
\caption[]{Comparison between CO and infrared modelling}
\begin{flushleft}
\begin{tabular}{llll|ll} \hline
Lum. & ${\kappa}_{60 \mu m}$ & ${\dot{M}}_{\rm CO}$ & 
${\Psi}_{\rm CO}$ & ${\dot{M}}_{\rm IR}$ & ${\Psi}_{\rm IR}$ \\
(\lsol) & cm$^2$/gr & $M_{\odot}/yr$&10$^{-2}$&$M_{\odot}/yr$ &10$^{-2}$ \\ \hline
 10~000  & 326 & 1.50$\times$ 10$^{-5}$& 0.125 & 1.10$\times$ 10$^{-5}$& 0.137 \\
 15~000  & 245 & 1.83$\times$ 10$^{-5}$& 0.167 & 1.73$\times$ 10$^{-5}$& 0.135 \\ 
 20~000  & 245 & 2.11$\times$ 10$^{-5}$& 0.167 & 2.37$\times$ 10$^{-5}$& 0.113 \\
 25~000  & 283 & 3.40$\times$ 10$^{-5}$& 0.100 & 3.02$\times$ 10$^{-5}$& 0.099 \\
 30~000  & 204 & 3.11$\times$ 10$^{-5}$& 0.167 & 3.67$\times$ 10$^{-5}$& 0.104 \\
\hline
\end{tabular}
\end{flushleft}
\end{table}

The uncertainty in the mass loss rate and dust-to-gas ratio derived
from the dust shell is about 8\% and 6\% due to uncertainties in the
assumed mass (varied to 0.6 \msol) and $v_0$ (varied to 0 \ks).  In
the ideal situation the mass loss rates and dust-to-gas ratios from
the two methods would be identical to indicate a perfectly consistent
model. The best overall agreement based on this analysis is achieved
for a luminosity of 15~000 \lsol.

Given all these considerations it is likely that the luminosity of
IRC~+10~216 is between 10~000 and 15~000 \lsol\ and its distance is
between 110 and 135 pc. The present-day mass loss rate is (1.5 $\pm$
0.3) $\times 10^{-5}$ \msolyr\ and the gas-to-dust ratio is about 700
$\pm$ 100. The dust opacity at 60 $\mu$m is found to be of order 250
cm$^2$gr$^{-1}$. The CO abundance is 1.1 $\times$ 10$^{-3}$ relative
to H$_2$.

Compared to the previously proposed models mentioned in the
introduction this represents a somewhat lower mass loss rate, partly
because these other models adopted larger distances as well in many
cases. The CO abundance is larger than in most previous models, but in
agreement with theoretical expectations. 

The gas-to-dust ratio we find is larger that the canonical value of
200 often quoted. On the other hand there have been other recent
studies that found larger gas-to-dust ratios for carbon stars as well,
e.g. a gas-to-dust ratio of 550 for IRC~+10~216 for a distance of 135
pc (Ivezi\'c \& Elitzur 1996), a value of 1230 for IRC~+10~216
(Winters et al. 1997b), a value of 530 for AFGL 3068
(Winters et al. 1997a). Note that in the model of Crosas \& Menten
(1997) the gas-to-dust ratio is fixed at 100, and they mention that a
larger value would result in lines that are consistently too low.


\acknowledgements{
Dr. Truong-Bach kindly made available the IRAM J = 1-0 and 2-1 data
described in Truong-Bach et al. (1991). 
We would like to thank John Howe, Mark Heyer, Linda Tacconi, Andy
Harris  and Reinhard Genzel, who were involved to some extent 
in the CO (6-5) observations.
This research has made use
of the SIMBAD database, operated at CDS, Strasbourg, France.
} 
vspace{-0.3mm} 

{}

\end{document}